\documentclass[aps,twocolumn,preprintnumbers,amsmath,amssymb,floatfix,groupedaddress,nofootinbib]{revtex4}

\usepackage{enumerate}
\usepackage{graphicx}
\usepackage{color}
\usepackage[utf8]{inputenc}
\usepackage{enumitem}
\usepackage{appendix}

\usepackage{fontenc}  
\usepackage{amsmath}

\usepackage[francais]{babel}
\newcommand{\beq}{\begin{equation}}
\newcommand{\eeq}{\end{equation}}
\newcommand{\bea}{\begin{eqnarray}}
\newcommand{\eea}{\end{eqnarray}}
\newcommand{\ba}{\begin{array}}
\newcommand{\ea}{\end{array}}

\newcommand{\bef}{\begin{figure}}
\newcommand{\eef}{\end{figure}}

\begin{document}

%%%% Article title to be placed here
\title{Revisiting Quantum Contextuality in an Algebraic Framework. }

\author{Mathias Van Den Bossche$^1$ and Philippe Grangier$^2$}

\affiliation{$^1$ Thales Alenia Space, 26, avenue J.-F. Champollion, 31037 Toulouse, France}
\affiliation{$^2$ Laboratoire Charles Fabry, IOGS, CNRS, 
Universit\'e Paris Saclay, F91127 Palaiseau, France.}

\begin{abstract}
Within the framework of quantum contextuality, we discuss the ideas of extracontextuality and extravalence, that allow one to relate Kochen-Specker's and Gleason's theorems. We emphasize that whereas Kochen-Specker's is essentially a no-go theorem, Gleason's provides a mathematical justification of Born's rule. Our extracontextual approach requires however a way to describe the ``Heisenberg cut''. Following an article by John von Neumann on infinite tensor products, this can be done by noticing that the usual formalism of quantum mechanics, associated with unitary equivalence of representations, stops working when countable infinities of particles (or degrees of freedom)  are encountered. This is because the dimension of the corresponding Hilbert space becomes  uncountably infinite, leading to the loss of unitary equivalence, and to sectorisation. Such an intrinsically contextual approach provides a unified mathematical model including both quantum and classical physics, that appear as required incommensurable facets in the description of nature.
\end{abstract}

\maketitle

\section{Introduction}
\vspace{-2mm}
The usual formalism of quantum physics makes use of separable Hilbert spaces{\color{black}\footnote{This means that there exists a countable basis in such spaces.} and operators acting in it, that form a type-I von Neumann algebra, according to the classification introduced in the 1930's by Murray and von Neumann \cite{FoP2023,JvN1939}.
In this article} we will argue that this mathematical framework is unduly restrictive, and that  it is 
worthwhile exploring what happens when the Hilbert space is non-separable and the operators acting on its 
vectors form a type-II or type-III von Neumann algebra. The motivation is that this extension appears as 
a reasonable framework to describe macroscopic systems with such mathematical tools, as explained below. 
However, operator algebras are unfriendly for non-mathematically oriented physicists, and they appear more as a subfield of mathematics than of physics. It addition,  it is not so obvious to find a clear-cut situation where type-I fails, and where higher-type operator algebras are required. 

In this paper, we want to show how such sophisticated mathematical objects can be very physically relevant to describe large systems, and may shine some light on foundational questions of quantum mechanics. Correspondingly, it will appear that the clear-cut situation where type-I fails is nothing but the so-called quantum measurement problem. 

In Section \ref{revisit} we will start from a simple discussion of contextuality in standard Quantum Mechanics (QM), and show how it relates to the whole structure of QM through Gleason's theorem and Born's rule. This leads however to a picture that cannot avoid including the ``Heisenberg cut'', separating (quantum) systems from (classical) contexts. In Section \ref{macroModel} we recall some basic results of \cite{FoP2023}, based on John von Neumann's  work on infinite tensor products \cite{JvN1939}, and we show how build up a complete and consistent picture of QM. 

\section{Quantum contextuality revisited.}
\label{revisit}

\subsection{From Kochen-Specker to Gleason.}

For the clarity of the presentation, we will introduce quantum contextuality in a simple way, based upon usual textbook quantum mechanics\footnote{Here we only consider 
 projective measurements (PVM), where, as explained in \cite{qc}, all the notions of compatibility, i.e., joint measurability, nondisturbance, outcome repeatability, etc., are equivalent to the commutativity of the observables. This is the original framework of the discussion on quantum contextuality, and though it has been extended in many other interesting directions  \cite{qc}, it is enough for our current purpose.}. Then we will open some questions, and propose a way to answer them. Finally, we will draw some conclusions from our approach. 

Within basic textbook quantum mechanics (QM), let us consider a complete quantum measurement of a quantity $A$, where ``complete'' means with no degeneracy left;  usually such a measurement is associated with a complete set of commuting operators (CSCO), that we summarize in a single ``observable'' operator  $\hat A$.   Denoting as $a_i$ and $| u_i \rangle$ the eigenvalues and eigenstates of $\hat A$, one has from the spectral theorem  $\hat A = \sum_i \; a_i \; | u_i \rangle \langle u_i |$  with standard Dirac's notations.  
In the following $\hat A$ will be identified with  a {\bf context}, that can be understood with different meanings: on the physical side the context corresponds to a macroscopic and operational device, able to measure the quantity $A$, whereas on the mathematical side it corresponds to the  observable $\hat A$, or also to the above set of rank-one orthogonal projectors $\{| u_i \rangle \langle u_i | \}$, where $i$ goes from 1 to $N$, the dimension of the Hilbert space. In usual QM these physical and mathematical objects correspond to each other, and we will come back to that in the last part of the paper.  {\color{black} We emphasize that in our approach 
QM is basically a non-classical probability theory, relative to events that happen to physical objects made of systems within contexts; in this framework the role of ``agents" is the same as in classical probability theory.}

According again to standard QM, a measurement of $\hat A$ will give one of the eigenvalues $a_i$, whereas other complete measurements $\hat B$, $\hat C$... associated with other contexts, will similarly give one of their eigenvalues $b_j$, $c_k$...  In each context the corresponding observable gets one value among $N$ possible ones, so one says that the value $a_i$ is assigned to the observable $\hat A$. 

Now, what is quantum contextuality, in its simplest definition ? It is the observation that in QM, {\bf it is impossible to assign simultaneously values  with certainty to all observables in all possible contexts, each one seen as the experimental background of a measurement.}  This simple observation clashes with classical physics, where  the possibility of  such an assignment is considered essential to express  the laws of nature. The formal proof of quantum contextuality is given by the Kochen-Specker (KS) theorem, showing a contradiction between QM and all models attributing non-contextual values to some (well chosen)  set of observables.  Such a result  is ``negative'', or in  other terms it is a no-go theorem, showing that a large class of non-contextual models (typically the models with non-contextual hidden variables) contradict QM predictions and experimental observations \cite{qc}.  
%\\

However, it does not tell much about what is actually happening in QM: the whole purpose of the KS discussion is to tell what QM is not, rather than to tell what QM is -- which is the real issue we are interested in.  

As a step in this direction, we may ask the question: if it is impossible to assign simultaneously values  with certainty to all observables in all possible contexts, how to make sense of what is being observed and measured ? There is an answer from observation: given a result in a context, one cannot in general specify results in other contexts, but one can specify the {\bf probabilities} of such results.  Even more interestingly, the assignment of probabilities to the value of all observables (that is, to  any given measurement result) turns out to be non-contextual! In other words, all measurement results within different contexts can be predicted simultaneously, albeit {\color{black} as} a probability rule, and not a certainty rule.

Contrary to the previous negative result, this is major asset: as we will show in more details below, it is the basis of Gleason's theorem, mathematically establishing  Born's rule. It is therefore a ``positive'' result, much more interesting and powerful than Kochen-Specker's theorem. On the mathematical side,  we note also that Kochen-Specker's theorem can be seen as a corollary of Gleason's one, whereas the reverse is not true \cite{qc}. 
It is also important to note that there is a special case for probabilities: if we restrict ourselves only to certainties (probabilities $p$ = 0 or $p$ = 1),  they are actually values (eigenvalues of projection operators), not probabilities (average values of projection operators). Certainties are therefore contextual, as it is not possible to assign truth values (0 or 1) to all measurement results, this is another useful way to see Kochen-Specker's theorem. But one can assign non-contextual probabilities as said above, with values given by Born's law or Gleason's theorem. 
\vskip 2mm

As a conclusion of this introduction, QM is both contextual (for the measurement results) and non contextual (for the probabilities assigned to these results). Though this sounds hardly comprehensible, everybody agrees on that, and this is the current status of quantum contextuality as presented in \cite{qc}. 
Our purpose here is to make sense of this situation, by introducing a few additional ideas and definitions. 

% \vspace{-2mm}
\subsection{What is quantum (extra)contextuality ? }

In this section we will introduce some new ideas, still within the scope of the usual textbook QM. Then at some point we will diverge, as it will be seen below. 
\vskip 2mm

First, let us introduce a distinction between a usual pure quantum state $| \psi \rangle$, described as  a vector in the relevant  Hilbert space {\color{black}$\cal H$} for the considered system, and the same vector  considered as an eigenstate of a complete measurement, therefore within a given context as introduced above.  We will define a {\bf modality} as the association of $| \psi \rangle$ and the context $\hat A$. By construction a  modality is certain and repeatable: the same $| \psi \rangle$, and the associated eigenvalue, will be found again and again as long as the same $\hat A$ is measured on the same system.  
\vskip 2mm

From this definition one has   
$| \psi  \rangle = | u_i  \rangle$, where $| u_i  \rangle$ is an eigenstate of $\hat A$, and the modality  ``$| \psi  \rangle \equiv  | u_i  \rangle$ considered as an eigenvector of $\hat A$'' will be denoted as $| \psi  \rangle_A$. It is clear that the same vector $| \psi  \rangle$ may appear as an eigenstate of many other observables (actually, an infinity), as soon as the dimension of $\cal H$ is at least 3.  Since the changes of context are continuous, both physically and mathematically, there is actually an infinity of such observables. So there will be many modalities, e.g. $| \psi  \rangle_A$, $| \psi  \rangle_B$, $| \psi  \rangle_C$, all associated with the same $| \psi  \rangle$: in the langage of usual QM, they {\it are} the same $| \psi  \rangle$; but here they are different modalities, because they belong to different sets of mutually exclusive outcomes. 

So now we start moving away from the usual  language: let us consider that the different modalities $| \psi  \rangle_A$, $| \psi  \rangle_B$, $| \psi  \rangle_C$  are indeed different, though they are all associated with the same vector $| \psi  \rangle$, or equivalently the same projector $| \psi  \rangle  \langle \psi |$ (projectors are actually more suitable for reasons that will appear below). Clearly this creates an equivalence class between modalities, that we will call an extravalence class, where all modalities are associated with the same $| \psi  \rangle$.  This means that certainty can be transferred between contexts, because $| \psi  \rangle$ is an eigenstate in all the corresponding contexts, and the associated physical phenomenon will be called extracontextuality.  
\vskip 2mm

So far so good, but does this help with the previous annoying statement that QM is both contextual for measurement results,  and non contextual for probabilities assignments ? To see  that consider the following statements 

(i) when making measurements, certainty and reproducibility are warranted for a modality in a given context,  and are also warranted for extravalent modalities when changing context. But they are not warranted beyond that, due to contextuality in the assignment of measurement results as established by the KS theorem. 

(ii) in other cases, when changing the context from $\hat A$ to $\hat  B$, the result is probabilistic and  given by Born's rule 
$| \langle \phi | \psi \rangle |^2$ for an initial modality $| \psi  \rangle_A$ and final one $| \phi  \rangle_B$. This probability depends only on the extravalence classes of the initial and final modalities, in agreement with the non-contextuality in the assignment of probabilities, %which is 
a basic hypothesis of Gleason's theorem. 
\vskip 2mm

So by attributing the vector $| \psi \rangle$ not to a ``quantum state", but to an extravalence class of modalities, one gets a more transparent picture of what QM is telling us: certainty and reproducibility do exist, not only within a context, but also {\color{black} accross contexts} within an extravalence class; when changing context and extravalence class,  probabilities are needed, but they are non-contextual because they connect extravalence classes, whatever the contexts are. 

This argument  tells the ``how", but not yet the ``why": what would be needed is to explain, or at least justify, the origins of statements (i) and (ii).  This can be done within the framework called CSM, for Contexts, Systems and Modalities \cite{csm1,trsa,csm4b,myst}, and we will briefly summarize here some elements taken from \cite{csm4b}.  In this framework systems within contexts are defined as basic objects,  and modalities are properties associated with certainties, in agreement with the first part of statement~(i). Its second part, i.e. the  contextuality of the assignment of measurement results, appears as a consequence of a quantization postulate: for a given system, the maximum number of mutually exclusive modalities is bounded to $M$;
this requires a probabilistic description when contexts and extravalence classes are being changed  \cite{csm4b}.  
\vskip 2 mm

To get the probability law (ii), one assumes that the probabilities when changing contexts depend only on the extravalence class, and one attributes a projector $| \psi  \rangle  \langle \psi |$ to each extravalence class. These are strong assumptions, but they do agree with empirical evidence, and are justified in detail  in \cite{csm4b}, based on induction. 
Then it is easy to check that all hypotheses for Gleason's theorem are satisfied, and as a consequence, that Born's rule is justified as the only acceptable probability law. 
This argument can be made even more convincing by using Uhlhorn's theorem in addition to Gleason's,
in order to associate unitarity transforms to context changes \cite{csm4c}.

We see now that the consequences drawn from extracontextuality are considerably stronger than the ones drawn from contextuality alone: we not only get no-go theorems, but we reach the correct probabilistic structure of QM, essentially by providing a physical justification of Gleason's and Uhlhorn's hypotheses.  This automatically ensures that  Kochen-Specker's theorem is true also, as a corollary of Gleason's, without the need of examining many different scenarii in many different dimensions  \cite{qc}.  

%\vspace{-2mm}
\subsection{Confronting the Heisenberg cut. }

%\vspace{-1mm}
A conclusion of the above is that
extracontextuality and subsequently extravalence tell ``how much non-contextual'' QM can be, given that it is neither non-contextual (as it would be the case in classical physics) nor fully contextual (then there would be no connection between different contexts, and thus no objective theory at all).  
\vskip 2mm
 
Another conclusion is that the usual state vector $| \psi  \rangle$ is incomplete indeed, not due to any ``hidden variable'', but because it gives access to an actual  physical modality $| \psi  \rangle_A$ {\bf only when the context {\color{black}$\hat A$} has been specified}.  Correspondingly, $| \psi  \rangle$ can be turned into a non-trivial ($ p \neq 0,1$) probability distribution only when considering a new measurement context, that does not admit $| \psi  \rangle$ as a modality; otherwise one has again $p$ = 0 or 1. This feature implies that $| \psi  \rangle$ is predictively incomplete  \cite{JJ}, and allows the violation of Bell's inequalities, without requiring any nonlocality at the elementary level \cite{inference}. 
 \vskip 2mm
 
On the other hand, the modalities $| \psi  \rangle_A$, $| \psi  \rangle_B$, $| \psi  \rangle_C$  are indeed complete, with deep reaching consequences on all the usual ``paradoxes" of QM. For instance,  in the framework of the Einstein-Bohr debate, it means that $| \psi  \rangle$ is incomplete indeed, as claimed by Einstein, Podolsky and Rosen, and that it must be completed by ``the very conditions that allow future predictions'' (i.e. by specifying the context, either $\hat A$, or $\hat  B$, or $\hat  C$), as claimed by Bohr \cite{debate}. 
\vskip 3mm

The above ideas emphasize that a well defined quantum property, i.e. a modality, belongs to a system within a context. This is a quite objective statement, but clearly the system and the context are not described in the same way.  A system with $M$ mutually exclusive modalities is associated with a Hilbert space $\cal H$ of dimension $M$, whereas the context (as a CSCO)  is associated with operators acting in $\cal H$.  Such operators are constructed  from macroscopic  data, such as orientations or positions of  the apparatus, and the whole construction makes clear that {\bf both} systems and contexts  are needed, and that there is no proper way to make one ``emerge'' from the other, although the consistent relationship between both will be explained below. 

This leads to the (in)famous question of the ``Heisenberg cut" between system and context and of the universality of the quantum description.
 Within the CSM framework there are two ways to answer this question~:
 
 -- The first and simplest one is to postulate that quantum objects are made of systems within contexts \cite{CO2002,csm1}. This fits quite well with usual textbook QM, and allows retrieving Born's rule from Gleason's theorem \cite{csm4b}, as explained above. The main benefit here is that the usual ``quantum paradoxes" just vanish: QM is certainly not classical, but it results from contextual quantization, which fits quite well within the usual physical realism \cite{CO2002,csm1,myst} - though not within {\bf classical} physical realism. 
 \\
 
 -- A second way is to include  both systems and contexts in the same description as incommensurable objects, along the lines introduced e.g. by von Neumann in \cite{JvN1939}. More details are given in \cite{FoP2023,completing}, but this approach requires to use algebraic tools that have not been considered in textbook QM until now. More technically, the operators used in textbook QM {\color{black} belong} to type-I algebras in the Murray-von Neumann classification, whereas the approach quoted here requires typically higher-order algebras, that are typically  type-III \cite{FoP2023,completing}.  These algebraic tools open a way to include both systems and contexts in a unified framework. {\color{black} The price to pay is to } ``manipulate infinities", which happens to be quite possible mathematically, even though not popular in physics nowadays \cite{FoP2023}.  We develop these ideas in the next section. 

\section{Modelling macroscopic quantum systems}
\label{macroModel}
\vskip 1 mm

This section assumes that the usual formalism derived from the contextuality postulates to describe microscopic quantum systems should in a certain manner still hold when considering macroscopic systems built out of a large number of microscopic quantum system. But one has to take into account some changes of the mathematical tools that they involve. This change, however, matches surprisingly well the properties of macroscopic systems.   

\subsection{Macroscopic limit modeling}
\label{macroLimit}
The basic assumption we make to extend the description of microscopic quantum systems to macroscopic systems is reminiscent from statistical physics. In this domain, it is assumed that a macroscopic {\color{black} object is well described} as a system where the number $N$ of particles it contains $N\rightarrow\infty$. This allows e.g. using the central limit theorem as exact, as well as obtaining non-analytical correlation functions in critical systems.

This assumption has key consequences in our case. As a matter of facts, it means that the composite quantum system that we will be considering has now an infinite number of elements, and thus that the Hilbert space to accomodate its states is the result of an infinite tensor product of elementary Hilbert spaces. Even in the simplest case of an infinite chain of spins 1/2 where the elementary Hilbert spaces have a finite dimension 2, the dimension of the macroscopic Hilbert state space is $2^{\aleph_0} = \aleph_1$, i.e. has the power of continuum. This means that there is no countable basis that spans this macroscopic Hilbert state space, i.e. it is non-separable. 
{\color{black}The operators acting in this space form in most cases a type-III von Neumann algebra, that  contains by construction no finite non-zero projection.  Type III is used in quantum field theory and statistical physics, for systems with an infinite number of particles or degrees of freedom \cite{completing}.} 

\subsection{Properties of macroscopic system state space}
\label{infTensProd}

\vskip 2 mm

The purpose of this section is a pedestrian introduction to the properties of the Hilbert spaces obtained by the approach suggested in the previous section, i.e. infinite tensor products of separable Hilbert spaces that result from the $N\rightarrow \infty$ limit. {\color{black} Following \cite{JvN1939}, we will first outline the meaning of the convergence of an infinite product of complex numbers, then outline the construction of a Hilbert space that results from an infinite tensor product, that will be denoted as ITP. Then we will give the most important properties of such a Hilbert space, and of the physically relevant operators acting in it. For a more detailed presentation we refer to \cite{FoP2023, JvN1939}.}

\vskip 2 mm

\noindent {\bf Infinite complex products} -- The basic predictions of quantum physics lie in probability amplitudes, i.e. scalar products between states 
in a Hilbert space. As scalar products of tensor products are products of elementary scalar products,
$(\langle \psi_1 \vert \otimes \langle \psi_2 \vert)(\vert \phi_1\rangle \otimes \vert \phi_2\rangle)=\langle \psi_1\vert\phi_1\rangle \langle \psi_2\vert\phi_2\rangle$, 
infinite tensor product require defining properly the existence of infinite products. 
Let $I$ be an infinite set of indices (countable or not), and let $\{ z_\alpha \}, \alpha \in I$ be a set 
of complex numbers labelled in $I$. The infinite product of the $z_\alpha$'s converges to complex value 
$Z$ iff one can get as close as desired to $Z$ with a finite partial product of selected 
$z_\alpha$'s with distinct $\alpha$'s.  One can show that this value is unique. In the case where $I$ is a countable ordered set, this is equivalent to the convergence of the series of the logarithm of the product to $\ln(Z)$. In terms of $\alpha$ values, it means that all but a finite number of the {\color{black}$z_\alpha$'s} accumulate in the vicinity of 1 in the complex plane. A notion of quasi-convergence can be introduced usefully, that corresponds to accumulation of $z_\alpha$'s in the vicinity of the unit circle, with no constraint on the argument.

\vskip 2 mm

\noindent {\bf Infinite tensor products (ITP)} -- Let us consider an infinite set of separable Hilbert spaces (i.e. with finite dimension or countably infinite dimension) $\{ \mathcal{H}_\alpha \}$. 
One defines a (quasi-) convergent sequence $\{\vert \psi_\alpha\rangle\}_{\alpha\in I}$ of vectors by selecting one 
$\psi_\alpha$ in each $\mathcal{H}_\alpha$ (assuming the axiom of choice) and requiring that $\prod_{\alpha \in  I} \langle\psi_\alpha\vert\psi_\alpha\rangle$ (resp. quasi-)converges as defined above. 
One has to distinguish the trivially convergent sequences (where at least one of the factors is 0) from the non-trivial convergent sequences on which the following is built. 
With these notions in mind, one can define linear forms on the set of convergent sequences, first by selecting another 
convergent sequence $\{\vert \phi_\alpha\rangle\}_{\alpha\in I}$ and forming 
$\phi[\{\vert \psi_\alpha\rangle\}]:=\prod_{\alpha \in  I} \langle\phi_\alpha\vert\psi_\alpha\rangle$, 
and then by expanding this to a linear combination of such linear forms with $N$ convergent sequences 
$\{\vert \phi^k_\alpha\rangle\}_{\alpha\in I,k=1,N}$ and $N$ complex coefficients $\Phi_k$ defining 
$\Phi[\{\vert \psi_\alpha\rangle\}]:=\sum_{k=1,N} \Phi_k^*\prod_{\alpha \in  I} \langle\phi^k_\alpha\vert\psi_\alpha\rangle$.
One can form the scalar product of such linear forms and show that their set is a Hilbert space. 
This scalar product allows defining a distance (the norm of the difference of vectors) and thus, taking $N\rightarrow\infty$, 
proceed with the topological completion of the set (just as the rational numbers are completed to build the real numbers). 
Then, the infinite tensor product of the  $\mathcal{H}_\alpha$'s is defined as the dual of this completion of the set of linear forms, noted $\mathcal{H}=\otimes_{\alpha\in I}\mathcal{H}_\alpha$. So, not only is it well defined but it can be shown to be a metrisable Hilbert space. {\color{black}Although this construction is rather lengthy,  it is rigorous and it shows that the language of quantum physics keeps existing with an infinite number of elements.}

\vskip 2 mm

\noindent {\bf Sectorisation} -- With the above definition, one can explore the properties of $\mathcal{H}$. 
One can define a relation between two vectors $\vert \Psi\rangle:= \otimes_{\alpha\in I}\vert \psi_\alpha\rangle $ and $\vert \Phi\rangle:= \otimes_{\alpha\in I}\vert \phi_\alpha\rangle $ in $\mathcal{H}$ as 
$\vert \Psi \rangle \sim \vert \Phi \rangle \Leftrightarrow \sum_{\alpha\in I}\vert \langle \phi_\alpha \vert \psi_\alpha\rangle - 1 \vert $ converges. 
One can show that this is an equivalence relation. Let us call `sectors' its equivalence classes.
It can be shown that when the above two vectors belong to the same sector, it means that $\vert \phi_\alpha\rangle \neq \vert \psi_\alpha \rangle$ for only a finite number of $\alpha$'s. 
This means that the space spanned by the vectors of one sector is of countable dimension, and that there are uncountably many sectors. 
It can be proven that in each sector there is at least one vector which all elements are normed, so is itself normed.
The most important property is that {\bf sectors are orthogonal} i.e. if $\vert \Psi \rangle$ and $\vert \Phi \rangle$ are in different sectors, then $\langle \Psi \vert \Phi \rangle = 0$ and if they are in the same sector, $\langle \Psi \vert \Phi \rangle = 0$ 
means that at least one  $\langle \psi_\alpha \vert \phi_\alpha \rangle = 0$. Beyond being orthogonal, their direct sum is $\mathcal{H}$. 
Finally, one can show that the sectorisation builds up progressively, in the sense that if 
$\vert \Psi \rangle$ and $\vert \Phi \rangle$ are not in the same sector, for any $\varepsilon > 0$ one can find a finite set $J\subset I$ of $N$ indices $\alpha$'s, all distinct, so as to build $\vert \Psi_N\rangle:= \otimes_{\alpha\in J}\vert \psi_\alpha\rangle $ and $\vert \Phi_N\rangle:= \otimes_{\alpha\in J}\vert \phi_\alpha\rangle$ such that 
$\vert \langle \Psi_N \vert \Phi_N \rangle \vert < \varepsilon$. The physical meaning of this results will be detailed in the next section. Before doing so, let us explain the main properties of operators that act in $\mathcal{H}$.

\vskip 2 mm

\noindent {\bf Composite, bounded  operators} -- Let us consider now the operators that are built from linear operators $\hat A_\alpha$'s that act in the elementary $\mathcal{H}_\alpha$'s. 
This requires first to define the completion of $\hat A_\alpha$ as $\tilde A_\alpha:= \hat A_\alpha \otimes (\otimes_{\beta\neq\alpha}\hat I_\beta)$ (where $\hat I_\beta$ is the identity in $\mathcal{H}_\beta$). 
One can show that for each $\alpha$, the bounded $\hat A_\alpha$'s form a ring $\mathcal{B}_\alpha$ and so the $\tilde A_\alpha$'s form a `completed' ring $\mathcal{\tilde B}_\alpha$ (i.e. stable under addition and multiplication). 
One can then merge all the $\mathcal{B}_\alpha$'s in a larger ring $\mathcal{B}^\#$ that is the ring of all bounded 
operators on $\mathcal{H}$ that can be built from elementary operators, and that is in general a sub-ring of the ring 
{\color{black} $\mathcal{B}$ }of all bounded operators on $\mathcal{H}$ (if $I$ is finite, $\mathcal{B}^\# = \mathcal{B}$). 
The most important property is that {\bf the operators of $\mathcal{B}^\#$  commute with the projectors on each sector}. In other words, operators built from elementary operators do not connect different sectors; {\color{black} however, more general operators connecting sectors do exist, but they cannot be formed ``from the bottom".} Finally, as for sectorisation, one can show that there is a gradual onset of this property.  Let $\hat A$ be a bounded operator in $\mathcal{B}^\#$. 
If  $\vert \Psi \rangle$ and $\vert \Phi \rangle$ are not in the same sector, for any $\varepsilon > 0$ 
one can find a finite set $J\subset I$ of $N$ indices $\alpha$'s, all distinct, so as to build 
$\vert \Psi_N\rangle:= \otimes_{\alpha\in J}\vert \psi_\alpha\rangle $, 
$\vert \Phi_N\rangle:= \otimes_{\alpha\in J}\vert \phi_\alpha\rangle$ and the restriction $\hat A_N$ of $\hat A$ to the $\otimes_{\alpha \in J}\mathcal{H}_\alpha$ such that
$\vert \langle \Psi_N \vert \hat A_N\vert\Phi_N \rangle \vert < \varepsilon$.  
Let us come now to the meaning these surprising mathematical properties have, if used as the language of quantum physics.

\subsection{Consequences on the properties of macroscopic systems}

\vskip 1 mm

Let us consider a quantum system made of $N$ degrees of freedom, each described by a microscopic or elementary Hilbert space $\mathcal{H}_\alpha, \alpha = 1,...,N$. The full Hibert space of the system is then $\mathcal{H}=\otimes_{\alpha=1}^N\mathcal{H}_\alpha$. As explained in section \ref{macroLimit}, we will consider that a macroscopic quantum system is well modelled, as in Statistical Physics, by the macroscopic limit $N\rightarrow \infty$. This means that the Hilbert space of this system shall have all the properties outlined in section \ref{infTensProd}. Let us expand on these properties now.

\vskip 2 mm

\begin{figure}[h]

%\begin{minipage}{22pc}
\includegraphics[width=\columnwidth]{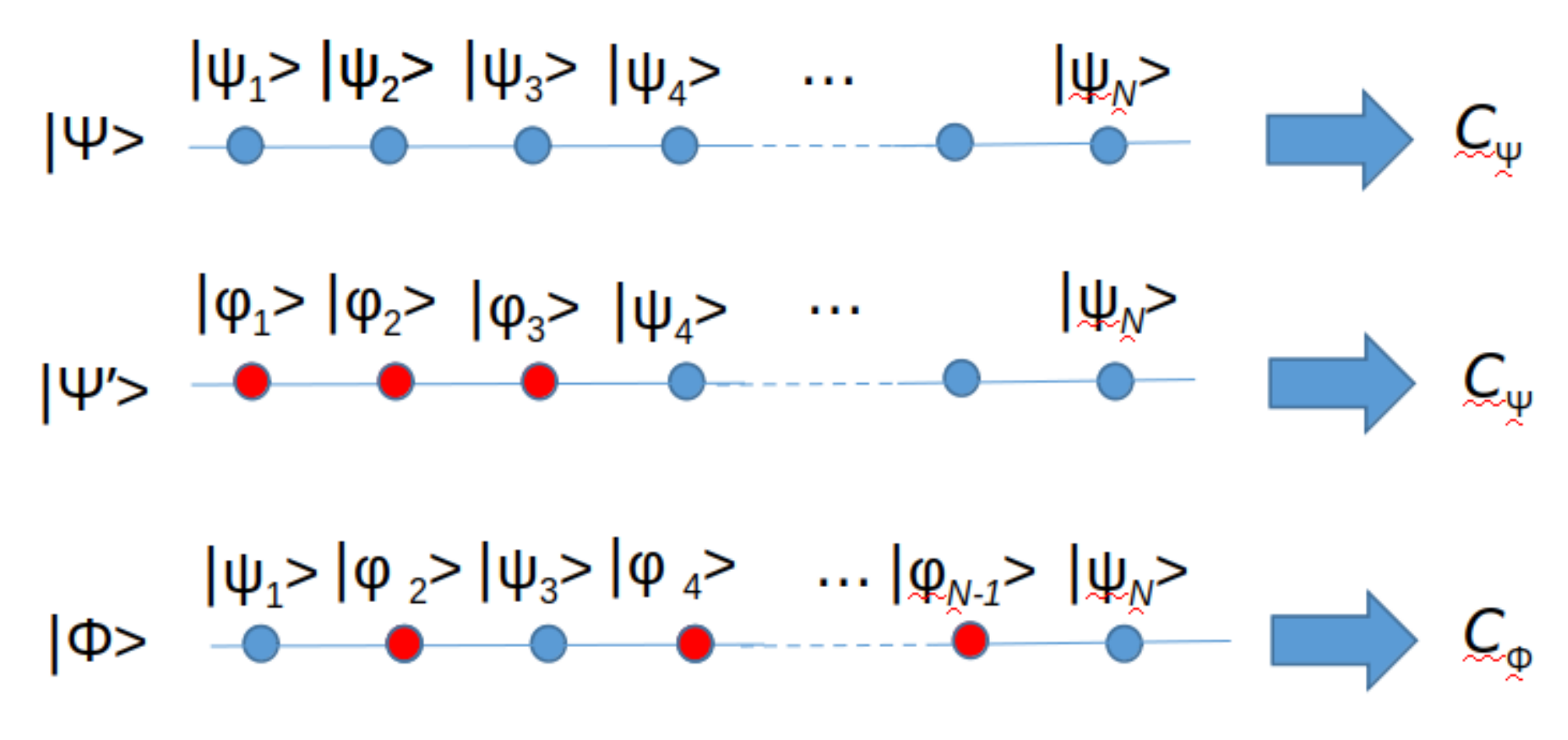}
\caption{\label{sectors}Illustration of sectors on a macroscopic chain.}
%\end{minipage}
\hspace{2pc}%
%\begin{minipage}{14pc}
\includegraphics[width=0.8 \columnwidth]{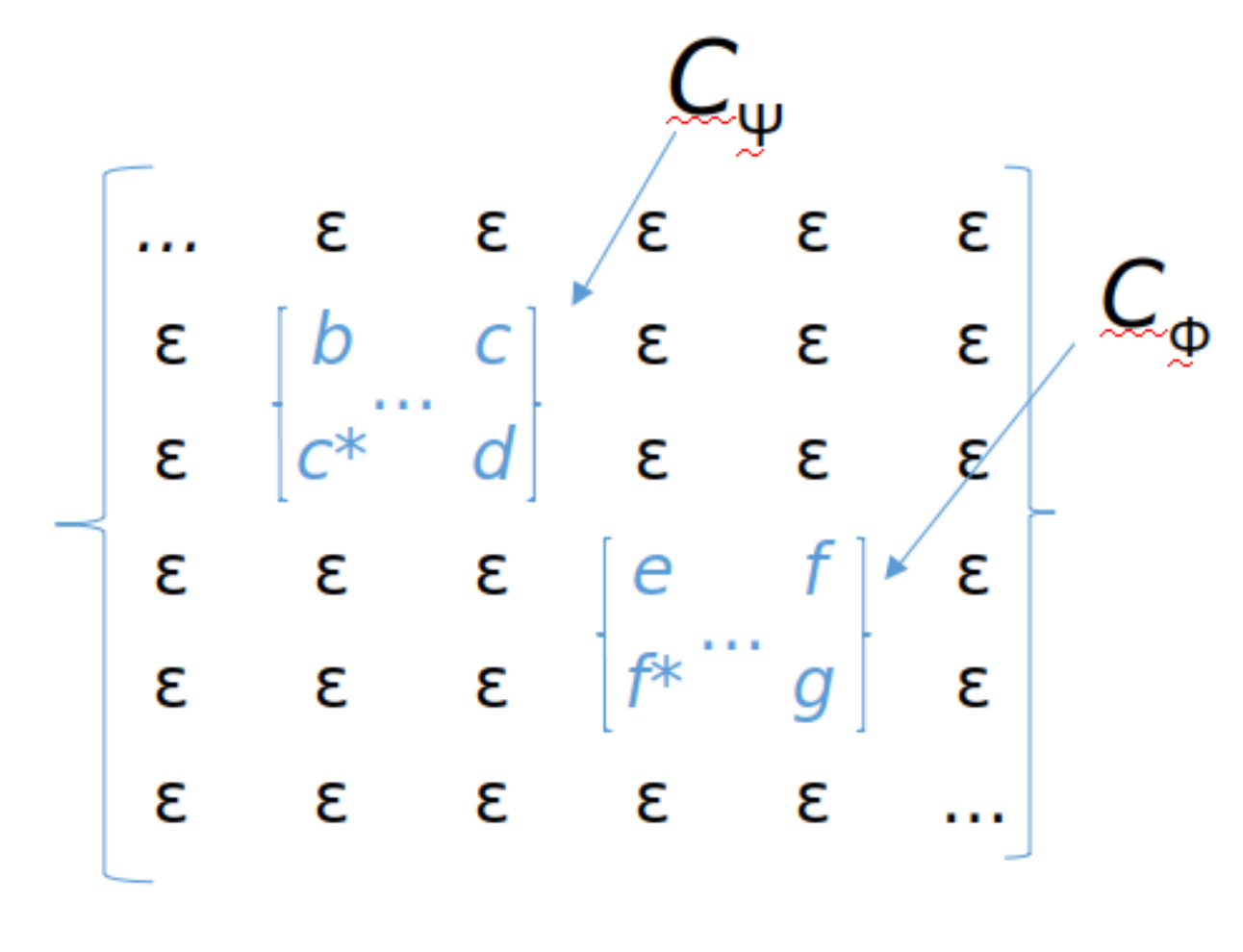}
\caption{\label{blockDiag} Typical shape of an operator built from elementary operators in an ITP.}
%\end{minipage}
\end{figure}

\noindent {\bf Sectorisation} is the first significant property of infinite tensor products. In this case, it means that $\mathcal{H}$ splits into uncountably many orthogonal subspaces $\{\mathcal{C}\}$ each of which is of countable dimension. 
Figure \ref{sectors} examplifies this with two-dimensional (blue or red) $\mathcal{H}_\alpha$'s. 
A the top, we have a given state $\vert \Psi\rangle $ that defines a sector $\mathcal{C}_\Psi$. 
In $\vert \Psi^\prime \rangle$ in the middle, a finite number of elementary states (namely, 3) have been changed from the first state. 
At the bottom, an infinite number of elementary states have been changed to build $\vert \Phi \rangle$  (namely the state of every odd element). 
One sees that there is a countable infinite way of changing a finite number of states in this system. 
It could be even finite if one decided to change only the state of a fixed finite set of elements. 
The subspace spanned by the states where only a finite number of elements are changed is thus separable. 
The second state $\vert \Psi^\prime \rangle$ thus remains in sector $\mathcal{C}_\Psi$.
One can say that even though this first two states are {\it microscopically} different, they are {\it macroscopically} 
similar (e.g. `blue').
Now considering the third state, an infinite number of states have been changed. 
By the above properties, this means that the bottom state $\vert \Phi \rangle$ is in a different sector $\mathcal{C}_\Phi$. 
It is thus orthogonal to the first two states, even if each changed elementary state is not orthogonal to its 
initial counterpart. For instance assuming the elements are spins $\frac{1}{2}$, and that 
$\vert \psi_\alpha\rangle=\vert \uparrow\rangle$ while 
$\vert \phi_\alpha\rangle = \vert + \rangle = (\vert \uparrow\rangle+\vert \downarrow\rangle)/\sqrt{2}$, 
then 
$\langle \psi_\alpha\vert \phi_\alpha\rangle \neq 0$, 
yet 
$\langle \Psi \vert \Phi\rangle = 2^{-N/4} \rightarrow 0$ 
in the macroscopic limit, so indeed $\langle \Phi \vert \Psi\rangle = 0$ .
One also sees that there is uncountably many ways to change an infinite number of elementary states in this way. 
There are thus uncountably many macroscopically different states.  

\vskip 2 mm

\noindent {\bf Self-decoherence } -- Let us now turn to the main property of bounded operators built from microscopic 
operators, which is that they commute with the projectors on each sector in the macroscopic limit. 
This means that such a bounded 
operator does not connect states from different sectors. This has two consequences. First, building an operator 
such as a Hamiltonian as the combination of elementary, microscopic operators will yield operators that have 
vanishing elements between sectors, or that can be taken to arbitrarily small values by combining enough elementary subsystems, see Fig. \ref{blockDiag}. This is also true of evolution operators built this way, that will thus not take a state from one sector to another. This means in particular that evolution cannot lead a state in one sector to a superposition
of states from different sectors. As a second example, the density operator of a large system will also have vanishing 
elements between sectors (or that can be taken to arbitrarily small values by combining enough elementary 
subsystems). This means that the larger the system, the smaller the correlations between sectors, so the 
macroscopic system behaves macroscopically as in a mixed state, with microscopic quantum correlations at most. 
This is reminiscent of Zeh-Zurek decoherence, apart from the fact that no assumption of information loss to the 
environment and partial trace to model it is needed. Decoherence appears thus to be an intrinsic property of large quantum systems.

\vskip 2 mm

\noindent {\bf Superposition of macroscopic system states} can be considered to be produced by two 
different mechanisms, evolution (as in Rabi oscillations) or measurement (at least QND) on a non-orthogonal state.
From the above considerations, one sees that macroscopic evolution from an initial state will not bring it 
to a superposed state of macroscopically different {\color{black} sectors}. Similarly, a projective measurement 
on a non-orthogonal state does not yield any superposed result (or rather, has a vanishing probability amplitude 
to succeed) because the projection of the state of one sector on the state of another sector vanishes. 
This means that the usual processes to produce superposed states fail for macroscopic systems. Consistently 
with the block-diagonality of density operator, there is no way to have macroscopically superposed states -- 
there are no such Sch\"odinger cats. As superposition is the way to code the measurement indeterminism in the 
quantum formalism, this means that {\bf macroscopic systems are deterministic}, while superpositions do make 
sense in the microscopic realm, that has a share of randomness.

\vskip 2 mm

\noindent {\bf Quantum measurements} can be considered {\color{black} as involving} a special case of 
macroscopic systems. As a matter of facts, in a typical experimental setup, we have a microscopic quantum 
system $S$ and a macroscopic measurement device $D$ which form a macroscopic system $S\cup D$. 
If the states of $S$ are spanned by the $M$ eigenstates {\color{black} $\vert u_k\rangle$} of some observable $\hat A$, the state of $S$ can be written as 
$\vert \psi\rangle =\sum_{k=1}^M \psi_k{\color{black} \vert u_k\rangle}$.
The measurement device is made so that starting from an initial 'neutral' state $\vert D_0\rangle$, it transitions to a 
macroscopic state  $\vert D_k\rangle$ that corresponds to state ${\color{black} \vert u_k\rangle}$ of the system. 
Usually, one assumes that first, a unitary evolution takes 
$S\cup D$ from $\vert in\rangle:= \sum_{k=1}^M \psi_k {\color{black} \vert u_k\rangle} \otimes \vert D_0\rangle$ 
to a superposed state 
$\vert \mathrm{out} \rangle:= \sum_{k=1}^M \psi_k {\color{black} \vert u_k\rangle} \otimes \vert D_k\rangle$
and second, that decoherence due to leak of entanglement {\color{black} to} the device, the laboratory and the full universe, makes the density operator $\vert \mathrm{out} \rangle \langle \mathrm{out}\vert$ becomes diagonal {\color{black} (modelled by tracing over the non-system dimensions)} and thus $S\cup D$ ends up in a mixed state.

\vskip 2 mm

The above considerations show that if modelling large systems with ITPs is correct, 
a new scenario has to be proposed for the quantum measurement process. A first pre-measurement step occurs where 
the system becomes entangled with ancilla degrees of freedom, which remain described with a separable Hilbert 
space. A second step then occurs where ancillae themselves become entangled with a number of secondary degrees 
of freedom (e.g. photons that probe the ancillae) that are selected so as to highlight their different states 
with a different behaviour.  This involves adding energy in the {\color{black} $S \cup D$} (e.g to accelerate photoelectrons) and triggers a macroscopic cascade where the number of quantum degrees of freedom in interaction increases exponentially and reaches a {\color{black}extremely large} number.  If we assume that the resulting very  large system is correctly modelled 
by infinite tensor products, it means that the various final states $\vert D_k\rangle$ of $D$ (that need to be 
macroscopically different), lie in distinct sectors.  

\vskip 2 mm

As a result, no unitary evolution will lead $S\cup D$ to a 
superposed state such as $\vert out\rangle$. On the contrary, as $S\cup D$ gets nearer to such a state, the 
off-diagonal terms of the density operator shall decrease steadily, and vanish before any such state is reached. 
At the end of the process, one of the $\vert D_k\rangle$ is {\color{black} probabilistically selected. The fact that this selection is random can  be seen as forced by the absence of a one-to-one map between the separable state space of $S$ and  the non-separable state space of $S \cup D$, so there is no determined ``way to go" for the evolution. 
This argument appears as  an instantiation in the infinite case of the CSM reasoning  \cite{csm4b},  showing that quantum randomness is a necessary consequence of contextual quantization. }

\subsection{Connecting back the results with the postulates}

\vskip 2 mm

The CSM approach allows building usual quantum mechanics thanks to a limited number of empirical, realist and objective  postulates and two {\color{black}key} theorems of linear algebra. It  can be summarised as follows.

\begin{itemize}[label=$\bullet$]
\item {\it Quantum systems} are physical objects that can be characterised by physical quantities 
which value is obtained through measurements.  

\item {\it Contextuality:} Measurements are done by chosing a context, i.e. what concrete quantities
are measured simultaneously. Once a context is chosen, a quantum system can provide only a limited amount 
($M\leq \aleph_0$) of different issues to the measurements.  
These issues are the modalities of the system within the selected context.

\item  {\it Contexts} themselves are continuously infinitely many, they are self-contextual 
and can provide as much information on themselves as desired.

\item  {\it Quantum randomness:} When changing the context, there is in general no certainty on what modality will be selected by a new measurement. Only probabilities can be predicted.  

\item  {\it Extravalence:} 
{\color{black} When the number of mutually exclusive modalities (later on, the dimension of the Hilbert space) is at least 3, certainty and repeatability can be transferred 
through modalities {\color{black} that belong} to infinitely many different contexts, defining an equivalence relation between modalitities called extravalence. The usual state vector $| \psi \rangle$ or projector  $| \psi \rangle \langle \psi |$ is associated with an extravalence class, not with a particular modality.}

\item  {\it Unitary transformations} are provided by Uhlhorn's theorem applied to the above objects. Let us assume that 
$M$ mutually orthogonal projectors are associated with the $M$ mutually exclusive modalities within a given context, and that the same projector is associated to all modalities in an extravalence class. For consistency, the $M$ projectors in a context should remain orthogonal under a map $\Gamma$ corresponding to a change of context. Then Uhlhorn's theorem proves  that $\Gamma$ should be unitary or anti-unitary, showing that the structure of the Hilbert space is a mathematical consequence of the physical behavior required for the modalities. 
  
\item  {\it Born's rule} is finally obtained by  applying Gleason's theorem to the above objects, to provide the value of the probability to obtain a certain modality in a given context.
\end{itemize}
Let us compare the results of the previous section to these basic elements.
\begin{itemize}[label=$\bullet$]
\item Macroscopic systems that differ by an infinite number of elementary, microscopic elements can be in an 
uncountable {\color{black}(i.e. continuous)} number of states. They cannot be superposed and thus are deterministic. In other words, they are 
classical and non-contextual. Contexts, as experimental devices, are such macroscopic systems, and what 
infinite tensor products tell is that they have precisely the properties postulated about them.
\item Macroscopic systems that differ by a finite number of elementary, microscopic elements can be in {\color{black} in a finite, or at most in a 
countably infinite number} of states. They are thus still contextual, and the finite part that differs can be seen as a 
quantum system. They are described by Hilbert space operators and Born's rule.
\end{itemize}
\begin{figure}[h]
%\begin{minipage}{22pc}
\begin{center}
\includegraphics[width=0.75 \columnwidth]{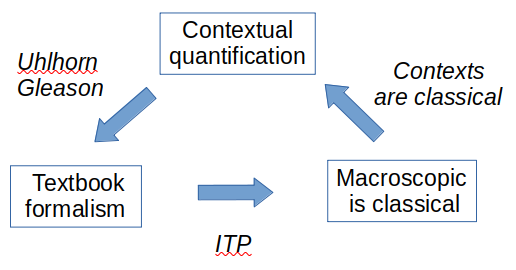}
\caption{\label{selfConsistency} Closing the loop of CSM. }
\end{center}
%\end{minipage}
\end{figure}
As a result, CSM, usual QM and ITP provide altogether a fully consistent conceptual setup to model both quantum 
systems and the classical systems they interact with{\color{black}, as illustrated in Fig.~\ref{selfConsistency}}. CSM has the interest of being based on not-so-abstract ideas
that can be used to develop an intuition on quantum systems, that is highly desirable to develop the emerging 
quantum technologies. This is in part thanks to its objective and realist basis. Finally, it leaves the much
exotic aspect of randomness at the heart of the approach, even if it provides justifications for it. We also
note that randomness seems to have a strong link with relativistic causality, and is thus all the more likely to 
be a fundamental aspect of our universe.

\vspace{- 2 mm}
\section{Conclusion}
\vspace{- 2 mm}
We have sketched a presentation of QM that relies on simple objective, realist and empirically understandable postulates. These postulates allow one to derive the usual textbook formalism up to dynamical equations. We push further the consequences by applying applying this formalism to the macroscopic limit that involves using type-III von Neumann algebras. The properties of these algebras allow understanding the behavior of large systems that fully support the initial assumptions. We highlight that the qualitative difference between quantum and classical system maps in this model on Cantor's hypothesis of continuum, i.e. that there is no other ``infinity" between the discrete one ($\aleph_0$) and the continuous one ($\aleph_1$). The non-unitary and genuinely random aspect of measurement occurs on the verge of this limit, that occurs physically when a macroscopic cascade maps a quantum phenomenon on a classical one.

\vskip 3mm

{\bf Acknowledgements.} 
PG thanks Franck  Lalo\"e, Roger Balian, Nayla Farouki, Alexia Auff\`eves, and  Ehtibar Dzhafarov for many discussions.

\end{document}